\documentclass[twocolumn,showpacs,preprintnumbers,amsmath,amssymb,nofootinbib,floatfix]{revtex4}

\usepackage{graphicx}

\def\Journal#1#2#3#4{{#1} {\bf #2}, #3 (#4)}
\def\ACTAPPB{Acta Phys. Polon. B}
 
\def\ANP{Ann. Phys.}
\def\APJ{Astrophys. J.}
\def\APJS{Astrophys. J. Suppl}

\def\EPJC{Eur. Phys. J. C}
\def\IJMPA{Int. J. Mod. Phys. A}

\def\JCAP{JCAP}

\def\JETPUSSR{JETP (USSR)}
\def\JPG{J. Phys. G}

\def\NJP{New. J. Phys.}
\def\NPB{Nucl. Phys. B}

\def\NPBSUPPL{Nucl. Phys. B. Proc. Suppl.}
\def\PLB{{Phys. Lett.} B}

\def\PLBOLD{Phys. Lett.}
\def\PAN{Phys. Atom. Nucl.}
\def\PRL{Phys. Rev. Lett.}
\def\PRD{Phys. Rev. D}

\def\PAN{Phys. Atom. Nucl.}
\def\PTP{Prog. Theor. Phys.}
\def\RMP{Rev. Mod. Phys.}
\def\SJNP{Sov. J. Nucl. Phys.}
\def\SCIENCE{Science}

\def\ZPC{Z. Phys. C}
\def\YFIZ{Yad. Fiz.}

\begin{document}

\title{Majorana CP phases in bi-pair neutrino mixing and leptogenesis}

\author{Teruyuki Kitabayashi}
\email{teruyuki@keyaki.cc.u-tokai.ac.jp}
\author{Masaki Yasu\`{e}}
\email{yasue@keyaki.cc.u-tokai.ac.jp}
\affiliation{Department of Physics, Tokai University, 4-1-1 Kitakaname, Hiratsuka, Kanagawa, 259-1292, Japan}
\date{\today}

\begin{abstract}
We estimate Majorana CP phases for a given flavor neutrino mass matrix ($M_\nu$) consistent with the bi-pair neutrino mixing, which 
is recently proposed to describe neutrino mixings given by $\sin\theta_{13}=0$ for the reactor neutrino mixing, 
$\sin^2\theta_{12} = 1-1/\sqrt{2}$ for the solar neutrino mixing and either $\sin^2\theta_{23} = \tan^2\theta_{12}$ 
or $\sin^2\theta_{23}=1-\tan^2\theta_{12}$ for the atmospheric neutrino mixing.  
Sizes of Majorana CP phases are evaluated so as to generate the observed baryon asymmetry in the universe via a leptogenesis scenario 
within the framework of the minimal seesaw model, where $M_\nu$  satisfies $\det ( M_\nu )=0$ and one active Majorana CP phase ($\phi$) is present. 
Assuming the normal mass hierarchy for light neutrinos and one zero texture for a $3 \times 2$ Dirac neutrino mass matrix, 
we find that $\phi$ lies in the region of $0.69 \lesssim \vert \phi\vert \lesssim 0.92$ [rad], which is converted into allowed regions of
$\alpha=\arg(M_{e\mu})$ and $\beta=\arg(M_{e\tau})$, where $M_{ij}$ ($i,j$=$e,\mu,\tau$) denote the $i$-$j$ matrix element of $M_\nu$. 
The phases $\alpha$ and $\beta$ turn out to satisfy $0.31 \lesssim \vert \alpha\vert \lesssim 0.40$ [rad] and $-1.25 \lesssim  \beta \lesssim -0.32$ [rad].
The approximate numerical equality of $\vert\phi\vert\approx 2\vert\alpha\vert$ is consistent with our theoretical estimation of $\phi=\phi_2-\phi_3 $ for 
$\phi_2=-(\alpha+\beta)$ and $\phi_3\approx \alpha-\beta$ valid for the normal mass hierarchy.
We also find the following scaling property:
$(M^\prime_{\mu\mu}-M^\prime_{ee}/t^2_{12})/M^\prime_{\mu\tau}=M^\prime_{\mu\tau}/(M^\prime_{\tau\tau}-M^\prime_{ee}/t^2_{12})=-M^\prime_{e\tau}/M^\prime_{e\mu}$
 ($t^2_{12}=\tan^2\theta_{12}=\sqrt{2}-1$),
where $M^\prime_{ij}$ stands for $M_{ij}$ evaluated on the basis of the Particle Data Group's phase convention.
 
\end{abstract}


\maketitle

\section{Introduction}
Experimental and theoretical studies of neutrino mixings have revealed various properties 
of neutrinos. The results from the atmospheric \cite{atmospheric}, solar \cite{solar}, reactor 
\cite{reactor, theta13} and accelerator \cite{accelerator} neutrino oscillation experiments have 
provided us with robust evidence that neutrinos have tiny masses and their flavor states are mixed with each other
\cite{Xing2010}. The squared mass differences of solar and atmospheric neutrinos, respectively,  defined by 
$\Delta m_{\odot}^2$ $\equiv$ $m_2^2-m_1^2$ and $\Delta m_{atm}^2$ $\equiv$ $m_3^2-m_1^2$, where $m_i$ 
$(i=1,2,3)$ is the mass of the corresponding generation of neutrinos, are observed to be \cite{Schwetz2008}:
\begin{eqnarray}
\Delta m_{\odot}^2 &=& 7.65_{-0.20}^{+0.23} \times 10^{-5} {\rm eV}^2, \nonumber \\
\left| \Delta m_{atm}^2 \right|  &=& 2.40_{-0.11}^{+0.12} \times 10^{-3} {\rm eV}^2.
\end{eqnarray}
The flavor mixing angles $\theta_{12}, \theta_{23}$ and 
$\theta_{13}$ are obtained as
\begin{eqnarray}
\sin^2 \theta_{12} &=& 0.304_{-0.016}^{+0.022}, \nonumber \\
\sin^2 \theta_{23} &=& 0.50_{-0.06}^{+0.07},      \nonumber \\
\sin^2 \theta_{13} &=& 0.01 _{-0.011}^{+0.016},
\label{Eq:nu-data}
\end{eqnarray}
where $\theta_{12}, \theta_{23}$ and $\theta_{13}$ stand for solar neutrino mixing angle, atmospheric 
neutrino mixing angle and reactor neutrino mixing angle, respectively. These mixing angles describe
the Pontecorvo-Maki-Nakagawa-Sakata (PMNS) matrix $U_{PMNS}$ \cite{UMNS} that converts mass eigenstates of neutrinos into 
flavor neutrinos.

One of the important and unsolved problem in neutrino physics is to understand CP properties of neutrinos. 
There are two sources of CP violations arising from Dirac CP phase and Majorana CP phase \cite{MajoranaPhase}.  
For three flavor neutrinos, 
CP violation is induced by one Dirac CP phase and two Majorana CP phases.  Since Dirac CP violation
involves the factor $\sin\theta_{13}$, no CP violation is induced by Dirac CP phase if $\sin\theta_{13}=0$.  
The current experimental data Eq.(\ref{Eq:nu-data}) are consistent with $\sin\theta_{13}=0$.
The latest data of $\sin\theta_{13}$ reported by T2K collaboration \cite{T2K} seem to suggest that 
$\sin\theta_{13}\neq 0$ at 90 $\%$ C.L, namely, $0.03 (0.04) < \sin^22\theta_{13} < 0.28 (0.34)$ 
for normal (inverted) mass hierarchy, giving $\sin^2\theta_{13}\gtrsim 0.0075$.  
The MINOS Collaboration has also reported to disfavor $\sin\theta_{13}=0$ \cite{MINOS}. 
Since these indications are not statistically sufficient, to get the 
definite confirmation of $\sin\theta_{13}\neq 0$ needs more data samples.  Theoretically, it is useful to 
construct a model giving $\sin\theta_{13}=0$, 
which can be regarded as a reference point to discuss effects of $\sin\theta_{13}\neq 0$.

If $\sin\theta_{13}=0$, Dirac CP phase is irrelevant. The remaining 
Majorana CP phases completely disappear from the oscillation probabilities and cannot be
measured by quite familiar oscillation experiments \cite{Mohapatra_Pal}. Although Majorana CP phases 
can enter in processes of neutrinoless double beta decay, 
the detection of Majorana CP violation has not been succeeded \cite{DoubleBeta}. 
On the other hand, in the leptogenesis scenario \cite{leptogenesisReviews}, the baryon-photon ratio in the universe ($\eta_B$)
is generated if Majorana CP phases exist and sizes of Majorana CP phases can be evaluated such that 
the observed ratio by WMAP collaboration \cite{WMAP} is reproduced.
There are theoretical discussions that predict $\sin\theta_{13}=0$ \cite{mu-tau, KitabayashiYasue, Tribimaximal, Scaling, Others}. 
We have recently proposed a bi-pair neutrino mixing scheme \cite{Bipair} that also
predicts $\sin\theta_{13}=0$ as well as $\sin^2\theta_{12} = 1-1/\sqrt{2}$ and either 
$\sin^2\theta_{23} = \tan^2\theta_{12}$ (referred to the case 1) or $\sin^2\theta_{23}=1-\tan^2\theta_{12}$ (referred to the case 2). 

In this paper, we would like to estimate sizes of phases of flavor neutrino masses associated with 
the bi-pair neutrino mixing.  To do so, we rely upon the seesaw mechanism \cite{Seesaw} to calculate $\eta_B$. 
Since $\eta_B$ depends on Majorana phases, to find constraints on phases of flavor neutrino masses, 
we have to derive direct relations between Majorana phases and phases of flavor neutrino masses.
It is convenient to adopt the minimal seesaw model \cite{minimalSeesaw,reconstruction}, where the number of physical phases associated with 
the seesaw mechanism are equal to that of CP phases of the flavor neutrino sector.

This paper is organized as follows. In the next section, we show a brief introduction to the bi-pair 
neutrino mixing and we also show phase structure of flavor neutrino masses leading to the bi-pair neutrino mixing. 
The direct relationship between Majorana phases and phases of flavor neutrino masses are derived. 
In section \ref{sec:Leptogenesis}, we give an outline of the minimal seesaw model and 
leptogenesis.  We perform numerical calculations to show the allowed region of Majorana phases and of 
phases of flavor neutrino masses. The last section is devoted to summary and discussions.

\section{Bi-pair neutrino mixing}\label{sec:BiPairMixing}
It is a good approximation that the reactor neutrino mixing angle is exactly zero \cite{theta13}. In 
this case, the PMNS matrix $U_{PMNS}$ given by the Particle Data Group \cite{PDG} to be 
$U_{PMNS}=U_0^{PDG}P^{PDG}$ : 
\begin{eqnarray}
U_0^{PDG}&=&\left(
  \begin{array}{ccc}
    c_{12}           & s_{12}          & 0 \\
    -c_{23}s_{12} & c_{23}c_{12}  & s_{23}  \\
    s_{23}s_{12}  & -s_{23}c_{12} & c_{23} 
  \end{array}
\right), \nonumber \\
P^{PDG}&=&\left(
  \begin{array}{ccc}
    1 & 0          & 0 \\
    0 & e^{i\phi_2/2}  & 0  \\
    0  & 0 & e^{i\phi_3/2} 
  \end{array}
\right),
\label{Eq:U0_P}
\end{eqnarray}
where $c_{ij}=\cos\theta_{ij}$, $s_{ij}=\sin\theta_{ij}$ $(i,j=1,2,3)$ and $\phi_{1,2}$ 
denote Majorana phases. 

\subsection{Texture}
The bi-pair neutrino mixing $U_{BP}$ is determined by a mixing matrix $U_0^{PDG}$ with two pairs of 
identical magnitudes of matrix elements. There are two possibilities of the bi-pair texture 
\cite{Bipair}, both of which predict $\sin^2\theta_{12} = 1 - 1/\sqrt{2} (\sim 0.293)$.

{\bf Case 1:}
The first possibility shows
\begin{eqnarray}
\vert \left( U_0^{PDG}\right)_{12} \vert &=& \vert \left( U_0^{PDG} \right)_{32} \vert, 
\nonumber \\
\vert \left( U_0^{PDG}\right)_{22} \vert &=& \vert \left( U_0^{PDG} \right)_{23} \vert.
\end{eqnarray}
These relations in turn provide useful relationship among the atmospheric neutrino mixing and the 
solar neutrino mixing as 
\begin{eqnarray}
\sin^2\theta_{23} = \tan^2\theta_{12}, \quad \tan^2\theta_{23} = \cos^2\theta_{12}. 
\end{eqnarray}
The bi-pair neutrino mixing in the case 1 is parameterized by only one mixing angle $\theta_{12}$: 
\begin{eqnarray}
U_{BP}=\left(
  \begin{array}{ccc}
    c_{12}           & s_{12}   & 0 \\
    -t_{12}^2      & t_{12}   & t_{12}  \\
    s_{12}t_{12} & -s_{12}  & t_{12}/c_{12} 
  \end{array}
\right), 
\end{eqnarray}
where $t_{ij}=\tan\theta_{ij}$ $(i,j=1,2,3)$. Numerically, the mixing angles are predicted to be:
\begin{eqnarray}
\sin^2\theta_{12} &=& 1 - \frac{1}{\sqrt{2}} \sim 0.293, \nonumber \\
\sin^2\theta_{23} &=& \tan^2\theta_{12} = \sqrt{2} - 1 \sim 0.414.
\end{eqnarray}
The bi-pair neutrino mixing well describes the observed solar neutrino mixing ($0.288 \le \sin^2\theta_{12} \le 0.326$);
 however, the atmospheric neutrino mixing is slightly inconsistent with the $1\sigma$ data 
($0.44 \le \sin^2 \theta_{23} \le 0.57$). It is expected that additional contribution to the atmospheric 
neutrino mixing angle is produced by the charged lepton correction
if a non-diagonal matrix element of charged lepton mass matrix only arises from a $\mu$-$\tau$ mixing mass
 so that $\theta_{23}$ can be shifted to the $1\sigma$ region without affecting the value of $\theta_{12,13}$.

{\bf Case 2:}
The second possibility shows
\begin{eqnarray}
\vert \left( U_0^{PDG} \right)_{12} \vert &=& \vert \left( U_0^{PDG} \right)_{22} \vert,
\nonumber \\
\vert \left( U_0^{PDG} \right)_{32} \vert &=& \vert \left( U_0^{PDG} \right)_{33} \vert.
\end{eqnarray}
The atmospheric neutrino mixing is related to the solar neutrino mixing as 
\begin{eqnarray}
\cos^2\theta_{23} = \tan^2\theta_{12}, \quad \tan^2\theta_{23} = 1/\cos^2\theta_{12}.
\end{eqnarray}
 The bi-pair neutrino mixing in the case 2 is parameterized by 
\begin{eqnarray}
U_{BP}=\left(
  \begin{array}{ccc}
     c_{12}            & s_{12}   & 0 \\
     -s_{12}t_{12}  & s_{12}   & t_{12}/c_{12}  \\
      t_{12}^2       & -t_{12}  & t_{12} 
  \end{array}
\right).
\end{eqnarray}
Numerically, the mixing angles are predicted to be:
\begin{eqnarray}
\sin^2\theta_{12} &=& 1 - \frac{1}{\sqrt{2}} \sim 0.293, \nonumber \\
\sin^2\theta_{23} &=& 1-\tan^2\theta_{12} = 2 - \sqrt{2} \sim 0.586.
\end{eqnarray}
Same as in the case 1, the atmospheric neutrino mixing is slightly inconsistent with the $1\sigma$ data. 

\subsection{General discussion on phase structure}
One may wonder what kind of flavor structure of a neutrino mass matrix $M_\nu$ is associated with the 
bi-pair neutrino mixing. To find phase structure of $M_\nu$ for the bi-pair neutrino mixing, we start 
our discussion with most general form of the PMNS mixing matrix $U_{PMNS} = U_0 P$ with
\begin{widetext}
\begin{eqnarray}
U_0 &=& 
  \left(
  \begin{array}{ccc}
     1 & 0 & 0 \\
     0 & e^{i\gamma} & 0 \\
     0 & 0 & e^{-i\gamma} 
  \end{array}
  \right)
  \left(
  \begin{array}{ccc}
     1 & 0 & 0 \\
     0 & c_{23} & s_{23} \\
     0 & -s_{23} & c_{23} 
  \end{array}
  \right)
  \left(
  \begin{array}{ccc}
     c_{13}                    & 0 & s_{13}e^{-i\delta} \\
     0                            &1  & 0 \\
     -s_{13}e^{i\delta}  & 0 & c_{13} 
  \end{array}
  \right)
  \left(
  \begin{array}{ccc}
     c_{12}                     & s_{12}e^{i\rho}  & 0 \\
     -s_{12}e^{-i\rho}  &c_{12}                    & 0 \\
     0                              & 0                          & 1 
  \end{array}
  \right)
  \nonumber \\
 &=&
  \left(
  \begin{array}{ccc}
     1 & 0 & 0 \\
     0 & e^{i\gamma} & 0 \\
     0 & 0 & e^{-i\gamma} 
  \end{array}
  \right)
  \left(
  \begin{array}{ccc}
     c_{12}c_{13}                                                                     & s_{12}c_{13}
e^{i\rho}                                                  & s_{13}e^{-i\delta}\\
     -c_{23}s_{12}e^{-i\rho}-s_{23}c_{12}s_{13}e^{i\delta}    & c_{23}c_{12}-s_{23}s_{12}s_{13}e^{i
(\rho+\delta)}  & s_{23}c_{13} \\
     s_{23}s_{12}e^{-i\rho}  -c_{23}c_{12}s_{13}e^{i\delta}   & -s_{23}c_{12}-c_{23}s_{12}s_{13}e^{i
(\rho+\delta)}  & c_{23}c_{13}
  \end{array}
  \right),
\end{eqnarray}
\end{widetext}
and
\begin{eqnarray}
P =
  \left(
  \begin{array}{ccc}
     e^{i\varphi_1}    & 0                      & 0 \\
     0                         & e^{i\varphi_2} & 0 \\
     0                         & 0                      & e^{i\varphi_3}
  \end{array}
  \right),
\end{eqnarray}
where $\gamma$, $\delta$, $\rho$ denote three Dirac phases and $\varphi_1$, $\varphi_2$, $\varphi_3$ denote 
three Majorana phases \cite{BabaAndYasue}. 

The phases $\gamma$ and $\rho$ in $U_0$ are redundant and we can remove these phases by the redefinition 
of flavor neutrinos resulting from the phase ambiguities present in  charged leptons. This redefinition 
can be express by a rotation matrix $R$ which has three phases $\theta_e$, $\theta_\mu$ and $\theta_\tau$:
\begin{eqnarray}
R =
  \left(
  \begin{array}{ccc}
     e^{i\theta_e}    & 0                       & 0 \\
     0                     & e^{i\theta_\mu} & 0 \\
     0                     & 0                      & e^{i\theta_\tau} 
  \end{array}
  \right).
\end{eqnarray}
After redundant phases are removed from $U_{PMNS}$, the mixing matrix becomes
\begin{eqnarray}
U_{PMNS}  \rightarrow U_{PMNS}^\prime = R U_{PMNS},
\end{eqnarray}
and accordingly the $3 \times 3$ symmetric flavor neutrino mass matrix 
\begin{eqnarray}
M_\nu =
  \left(
  \begin{array}{ccc}
     M_{ee} & M_{e\mu}      & M_{e\tau}      \\
               & M_{\mu\mu}  & M_{\mu\tau} \\
                &                        & M_{\tau\tau} 
  \end{array}
  \right),
\label{Eq:M_nu-original}
\end{eqnarray}
is shifted as follows:
\begin{eqnarray}
M_\nu   \rightarrow M_\nu^\prime = R^\dagger M_\nu R.
\label{Eq:M_nu}
\end{eqnarray}
For example, to obtain the Particle Data Group's form of the mixing matrix, we take $\theta_e = -\rho$,
 $\theta_\mu = -\gamma$ and $\theta_\tau = \gamma$, which give $R$ for the PDG version, $R_{PDG}$:
\begin{eqnarray}
R_{PDG} =
  \left(
  \begin{array}{ccc}
     e^{-i\rho}    & 0                    & 0 \\
     0                & e^{-i\gamma} & 0 \\
     0                     & 0              & e^{i\gamma} 
  \end{array}
  \right).
\end{eqnarray}
We, then, find the PDG version of $U_0$, $U_0^{\prime PDG}$:
\begin{widetext}
\begin{eqnarray}
U_0^{\prime PDG} =
  \left(
  \begin{array}{ccc}
     c_{12}c_{13}                                                                & s_{12}c_{13}
                                                            & s_{13}e^{-i\delta_{CP}}\\
     -c_{23}s_{12 }-s_{23}c_{12}s_{13}e^{i\delta_{CP}}    & c_{23}c_{12}-s_{23}s_{12}s_{13}e^{i\delta_
{CP}}  & s_{23}c_{13} \\
     s_{23}s_{12}  -c_{23}c_{12}s_{13}e^{i\delta_{CP}}   & -s_{23}c_{12}-c_{23}s_{12}s_{13}e^{i\delta_
{CP}}  & c_{23}c_{13} \\
  \end{array}
  \right),
\end{eqnarray}
\end{widetext}
and the PDG version of $P$, $P^{\prime PDG}$:
\begin{eqnarray}
P^{\prime PDG} =
  \left(
  \begin{array}{ccc}
     e^{i\phi^\prime_1}   & 0                    & 0 \\
     0                     & e^{i\phi^\prime_2} & 0 \\
     0                     & 0                   & e^{i\phi^\prime_3} 
  \end{array}
  \right),
\end{eqnarray}
 where 
\begin{eqnarray}
\delta_{CP} = \delta + \rho, \quad
\phi^\prime_1 = \varphi_1 -\rho, \quad 
\phi^\prime_{2,3} = \varphi_{2,3}.
\end{eqnarray}
After appropriate redefinition of Majorana phases, Eq.(\ref{Eq:U0_P}) turns out to be:
\begin{eqnarray}
U^{PDG}_{PMNS}=U_0^{PDG}P^{PDG},
 \label{Eq:U-PDG-PMNS}
\end{eqnarray}
where 
\begin{eqnarray}
\phi_2&=&2(\phi^\prime_2-\phi^\prime_1)
\nonumber\\
\phi_3&=&2(\phi^\prime_3-\phi^\prime_1).
 \label{Eq:phi-phiprime}
\end{eqnarray}
 The flavor neutrino mass matrix Eq.(\ref{Eq:M_nu-original}) is shifted to
\begin{eqnarray}
&&M_\nu^{PDG}  =
  \left(
  \begin{array}{ccc}
     M^\prime_{ee}  & M^\prime_{e\mu} & M^\prime_{e\tau}    \\
                                   & M^\prime_{\mu\mu}      & M^\prime_{\mu\tau}           \\
                                   &                                                    & M^\prime_{\tau\tau} 
  \end{array}
  \right)
\nonumber\\
&&\quad
=
  \left(
  \begin{array}{ccc}
     e^{2i\rho}M_{ee}  & e^{i(\rho+\gamma)}M_{e\mu} & e^{i(\rho-\gamma)}M_{e\tau}    \\
                                   & e^{2i\gamma}M_{\mu\mu}      & M_{\mu\tau}           \\
                                   &                                                    & e^{-2i\gamma}
M_{\tau\tau} 
  \end{array}
  \right).
  \label{Eq:MnuPDG}
\end{eqnarray}
%

\subsection{Phase structure in the bi-pair neutrino mixing}
For any type of neutrino mixings that give $\sin\theta_{13}=0$, 
it can be argued that Eq.(\ref{Eq:M_nu-original}) becomes the following mass matrix  \cite{YudaAndYasue}:
\begin{eqnarray}
M_\nu \vert_{\theta_{13}=0}  =
  \left(
  \begin{array}{ccc}
     M_{ee}  & e^{i\alpha} \vert M_{e\mu} \vert  & -t_{23}e^{i\beta} \vert M_{e\mu} \vert \\
                                                          & M_{\mu\mu}              & 
M_{\mu\tau}\\
                                                          &                                                       
& M_{\tau\tau} \\
  \end{array}
  \right),
  \label{Eq:Mnu_sin_theta13_0}
\end{eqnarray}
with
\begin{eqnarray}
M_{\tau\tau}   = e^{4i\gamma} M_{\mu\mu} + \frac{1-t_{23}^2}{t_{23}}e^{2i\gamma} M_{\mu\tau},
\label{Eq:M-tau-tau}
\end{eqnarray}
and 
\begin{eqnarray}
\gamma = \frac{\beta - \alpha}{2}.
\label{Eq:phase_gamma}
\end{eqnarray} 
The neutrino masses $m_1$, $m_2$, $m_3$ defined by
\begin{eqnarray}
U_{PMNS}^T M_\nu\vert_{\theta_{13}=0} U_{PMNS}
=
\left(
  \begin{array}{ccc}
     m_1 & 0      & 0 \\
     0      & m_2 & 0 \\
     0      & 0      & m_3 
  \end{array}
  \right),
\end{eqnarray}
are calculated to be
\begin{eqnarray}
m_1e^{-i\phi_1} &=& e^{2i\rho} M_{ee} - t_{12} \frac{e^{i\xi}\vert M_{e\mu} 
\vert}{c_{23}}, \nonumber \\
m_2e^{-i\phi_2} &=& e^{2i\rho} M_{ee} + \frac{1}{t_{12}} \frac{e^{i\xi}\vert 
M_{e\mu} \vert}{c_{23}}, \nonumber \\
m_3e^{-i\phi_3} &=& e^{2i\gamma}  M_{\mu\mu} +\frac{1}{t_{23}} M_{\mu\tau},
\label{Eq:m1m2m3}
\end{eqnarray}
where
\begin{eqnarray}
 \xi = \rho + (\alpha + \beta)/2.
 \label{Eq:xi}
\end{eqnarray}
The mixing angle $\theta_{12}$ is given by 
\begin{eqnarray}
\tan2\theta_{12} = \frac{2e^{i\xi} c_{23}^{-1} \vert M_{e\mu}\vert}{e^{2i\gamma}M_{\mu\mu} 
- t_{23}M_{\mu\tau} - e^{2i\rho}M_{ee }}.
\label{theta-12}
\end{eqnarray}
From Eq.(\ref{Eq:Mnu_sin_theta13_0}), we obtain $M_\nu^{PDG} \vert_{\theta_{13}=0}$:
\begin{eqnarray}
&&
M_\nu^{PDG} \vert_{\theta_{13}=0}  
\nonumber\\
&&\quad
=
  \left(
  \begin{array}{ccc}
     e^{2i\rho}M_{ee}  & e^{i\xi} \vert M_{e\mu} \vert  & -t_{23}e^{i\xi} \vert M_{e\mu} \vert \\
                                                          &e^{2i\gamma}M_{\mu\mu}              & 
M_{\mu\tau}\\
                                                          &                                                       
& e^{-2i\gamma}M_{\tau\tau} \\
  \end{array}
  \right).
\nonumber\\
\label{Eq:MnuPDG_sin_theta13_0-PDG}
\end{eqnarray}
Furthermore, by using (\ref{theta-12}) to eliminate $e^{2i\gamma}M_{\mu\mu}$, 
we find the following flavor structure:
\begin{eqnarray}
&&M_\nu^{PDG} \vert_{\theta_{13}=0}
= 
{e^{2i\rho }}{{M_{ee}}}I
\nonumber\\
&&\quad
+\left( {\begin{array}{*{20}{c}}
0&1&{ - {t_{23}}}\\
{}&{\frac{2}{{\tan 2{\theta _{12}}}}\frac{1}{{{c_{23}}}}}&0\\
{}&{}&{\frac{2}{{\tan 2{\theta _{12}}}}\frac{1}{{{c_{23}}}}}
\end{array}} \right){e^{i\xi }}\left| {{M_{e\mu }}} \right|
\nonumber\\
&&\quad
+\left( {\begin{array}{*{20}{c}}
0&0&0\\
{}&{{t_{23}}}&1\\
{}&{}&{\frac{1}{{{t_{23}}}}}
\end{array}} \right){M_{\mu \tau }}.
\label{Eq:M-nu-PDG}
\end{eqnarray}
For the bi-pair mixing scheme, the mixing angles in Eq.(\ref{Eq:M-nu-PDG}) are fixed to be 
$\tan^22\theta_{12}=2(\sqrt{2}-1)$ together with $\tan^2\theta_{23}=1/\sqrt{2}$ for the case 1 
and $\tan^2\theta_{23}=\sqrt{2}$ for the case 2.  A more transparent form of its flavor structure
can be obtained when either $m_1$ or $m_3$ vanishes as in the minimal seesaw model to be discussed in the next section.

Let us consider neutrinos exhibiting $m_1 = 0$, which corresponds to the normal mass hierarchy. Since there is the phase ambiguity in the 
charged lepton sector, we can choose three phases associated with flavor neutrino masses to be any values.  One may 
assign a specific value to $M_{\mu\tau}$ to be consistent with $m_1=0$ and take $M_{ee}$ and $M_{\mu\mu}$ to be real.
Accordingly, Eqs.(\ref{Eq:MnuPDG_sin_theta13_0-PDG}), (\ref{Eq:M-tau-tau}), (\ref{Eq:m1m2m3}) and (\ref{theta-12})
turn out to be
\begin{eqnarray}
&&
M^{PDG}_\nu \vert^{m_1=0}_{\theta_{13}=0}
\nonumber\\
&&\quad
 =
  \left(
  \begin{array}{ccc}
     e^{2i\rho}\kappa_e\left|M_{ee}\right|  & e^{i\xi} \vert M_{e\mu} \vert  & -t_{23}e^{i\xi} \vert M_{e\mu} \vert \\
                                                          &e^{2i\gamma}\left|M_{\mu\mu}\right|              & 
M_{\mu\tau}\\
                                                          &                                                       
& e^{-2i\gamma}M_{\tau\tau} \\
  \end{array}
  \right),
\nonumber\\
\label{Eq:Mnu_sin_theta13_0-real}
\\
&&
M_{\tau\tau} = e^{4i\gamma}\left| M_{\mu\mu}\right| + \frac{1-t_{23}^2}{t_{23}}e^{2i\gamma} M_{\mu\tau},
\\
&&
m_1e^{-i\phi_1} = e^{2i\rho} \kappa_e\left|M_{ee}\right| - t_{12} \frac{e^{i\xi}\vert M_{e\mu} 
\vert}{c_{23}}(=0), 
\nonumber\\
&&
m_2e^{-i\phi_2} = e^{2i\rho} \kappa_e\left|M_{ee}\right| + \frac{1}{t_{12}} \frac{e^{i\xi}\vert 
M_{e\mu} \vert}{c_{23}}, 
\nonumber\\
&&
m_3e^{-i\phi_3} = e^{2i\gamma}  \left|M_{\mu\mu}\right|  +\frac{1}{t_{23}} M_{\mu\tau},
\label{Eq:m1m2m3-real}
\\
&&
\tan2\theta_{12} = \frac{2e^{i\xi} c_{23}^{-1} \vert M_{e\mu}\vert}{e^{2i\gamma}\left|M_{\mu\mu}\right| 
- t_{23}M_{\mu\tau} - e^{2i\rho}\kappa_e\left|M_{ee}\right|},
\label{theta-12-real}
\end{eqnarray}
where the sign of $M_{ee}$ is taken care of by $\kappa_e=\pm 1$ for $-\pi/2\leq 2\rho \leq \pi/2$.
From $m_1=0$ in Eq.(\ref{Eq:m1m2m3-real}), we find  $\xi=2\rho$ leading to 
\begin{eqnarray}
\xi=2\rho=\alpha+\beta.
\label{Eq:xi=alpha+beta}
\end{eqnarray}
We then find that, from (\ref{theta-12-real}), $M_{\mu\tau}$ is given by
\begin{eqnarray}
{M_{\mu \tau }} = \frac{1}{{{t_{23}}}}\left( {{e^{i\left( {\beta  - \alpha } \right)}}\left| {{M_{\mu \mu }}} \right| - \frac{{{e^{i\left( {\alpha  + \beta } \right)}}\left| {{M_{e\mu }}} \right|}}{{{c_{23}}{t_{12}}}}} \right).
\label{Eq:phase-Mmutau}
\end{eqnarray}
The phase of $M_{\mu \tau }$ should be adjusted to satisfy Eq.(\ref{Eq:phase-Mmutau}).
The neutrino masses turn out to be:
\begin{eqnarray}
{m_2}{e^{ - i{\phi _2}}}& =& \frac{{{e^{i\left( {\alpha  + \beta } \right)}}\left| {{M_{e\mu }}} \right|}}{{{c_{12}}{s_{12}}{c_{23}}}},
\nonumber\\
{m_3}{e^{ - i{\phi _3}}}& =& \frac{1}{{s_{23}^2}}\left( {{e^{i\left( {\beta  - \alpha } \right)}}\left| {{M_{\mu \mu }}} \right| - \frac{{{e^{i\left( {\alpha  + \beta } \right)}}{c_{23}}\left| {{M_{e\mu }}} \right|}}{{{t_{12}}}}} \right).
\nonumber\\
\label{Eq:m_2-m_3}
\end{eqnarray}
The Majorana phase $\phi_2$ is simply given by
\begin{eqnarray}
\phi_2 = -(\alpha+\beta).
\label{Eq:phase_phi2}
\end{eqnarray}
The phase $\beta-\alpha$ can also be calculated from the equivalent relation of 
\begin{eqnarray}
\arg \left( {\beta  - \alpha } \right) = \arg \left( {t_{23}^2{m_3}{e^{ - i{\phi _3}}} + c_{12}^2{m_2}{e^{ - i{\phi _2}}}} \right).
\end{eqnarray}
The condition of $m^2_3\gg m^2_2$ leads to
\begin{eqnarray}
\phi_3 \approx \alpha-\beta.
\label{Eq:phase_phi3}
\end{eqnarray}
For the rest of paper, we use $\phi$:
\begin{eqnarray}
\phi = \phi_2 - \phi_3,
\label{Eq:CP-phi}
\end{eqnarray}
to denote Majorana CP phase
 
For the bi-pair neutrino mixing, the flavor neutrino mass matrix Eq.(\ref{Eq:M-nu-PDG}) is converted to be
\begin{widetext}
\begin{eqnarray}
M_\nu^{PDG} \vert_{\theta_{13}=0}^{m_1=0} &=&
\left(
  \begin{array}{ccc}
    A e^{i(\alpha + \beta)} \vert M_{e\mu} \vert  
           & e^{i(\alpha + \beta)} \vert M_{e\mu} \vert     
           & -\frac{A}{B} e^{i(\alpha + \beta)} \vert M_{e\mu} \vert 
    \\
           & \frac{A}{t_{12}^2}e^{i(\alpha+\beta)} \vert M_{e\mu} \vert + \frac{A}{B} M_{\mu\tau}  
           & M_{\mu\tau}  
           \\
           &        
           & \frac{A}{t_{12}^2}e^{i(\alpha+\beta)} \vert M_{e\mu} \vert + \frac{B}{A}M_{\mu\tau}
  \end{array}
 \right),
 \label{Eq:M_case1_case2}
 \end{eqnarray}
\end{widetext}
where $(A, B) = (c_{12}, 1)$ for the case 1 while $(A, B) = (1, c_{12})$ for the case 2.
It should be mentioned that the mass matrix $M_\nu^{PDG} \vert_{\theta_{13}=0}^{m_1=0}$ exhibits the following scaling property:
\begin{eqnarray}
&&\frac{M^\prime_{ee}}{M^\prime_{e\mu}}=A, 
\nonumber\\
&&\frac{M^\prime_{\mu\mu}-\frac{M^\prime_{ee}}{t^2_{12}}}{M^\prime_{\mu\tau}}=\frac{M^\prime_{\mu\tau}}{M^\prime_{\tau\tau}-\frac{M^\prime_{ee}}{t^2_{12}}}=-\frac{M^\prime_{e\tau}}{M^\prime_{e\mu}}=\frac{A}{B},
\label{Eq:M-ratios}
\end{eqnarray}
where $M^\prime_{ij}$ ($i,j$=$e$, $\mu$, $\tau$) stand for the matrix elements of $M_\nu^{PDG}$ as in Eq.(\ref{Eq:MnuPDG}).

\section{Leptogenesis}\label{sec:Leptogenesis}
In this section, first, we give an outline of the minimal seesaw model and leptogenesis, then, we 
show the allowed region of Majorana phases  from a numerical calculation.

\subsection{Minimal seesaw model}
In the minimal seesaw model \cite{minimalSeesaw}, we introduce two heavy neutrinos $N_1$ and $N_2$ 
into the standard model. We obtain a symmetric $3 \times 3$ light neutrino mass matrix by the relation 
of $M_\nu = -m_D M_R^{-1} m_D^T$, where $m_D$ is a $3 \times 2$ Dirac neutrino mass matrix and $M_R$ 
is a $2 \times 2$ heavy neutrino mass matrix. We assume that the mass matrix of the heavy neutrinos as 
well as of the charged leptons is diagonal and real. For the heavy neutrinos, $M_R$ takes the form of
\begin{eqnarray}
M_R = \left(
  \begin{array}{cc}
  M_1 & 0   \\ 
  0       & M_2
  \end{array}
\right)\quad (M_2>M_1).
\end{eqnarray}
The Dirac neutrino mass matrix $m_D$ can be expressed in terms of 6 parameters 
$a_1,a_2,a_3,b_1,b_2,b_3$ and two heavy neutrino masses $M_1,M_2$ as \cite{reconstruction}
\begin{eqnarray}
m_D = 
\left(
  \begin{array}{cc}
    \sqrt{M_1}a_1  & \sqrt{M_2}b_1   \\
    \sqrt{M_1}a_2  & \sqrt{M_2}b_2   \\
    \sqrt{M_1}a_3  & \sqrt{M_2}b_3   \\
  \end{array}
\right),
\end{eqnarray}
where one zero texture is assumed and
the light neutrino mass matrix is obtained from $M^{PDG}_\nu = -m_D M_R^{-1} m_D^T$ as
\begin{eqnarray}
&&M^{PDG}_\nu =
\left(
  \begin{array}{ccc}
    M^\prime_{ee} & M^\prime_{e\mu}      & M^\prime_{e\tau} \\
                  & M^\prime_{\mu\mu} & M^\prime_{\mu\tau} \\
                   &                         & M^\prime_{\tau\tau} \\
  \end{array}
\right)
\nonumber\\
&&\quad
=
\left(
  \begin{array}{ccc}
    a_1^2 + b_1^2   & a_1a_2 + b_1b_2   &  a_1a_3 + b_1b_3  \\
                               & a_2^2 + b_2^2     &  a_2a_3 + b_2b_3  \\
                               &                   &  a_3^2 + b_3^2   \\
  \end{array}
\right),
\label{Eq:Mnu}
\end{eqnarray}
whose phase structure is determined by Eq.(\ref{Eq:MnuPDG}), which is described by Eq.(\ref{Eq:M_case1_case2}).
The condition Eq.(\ref{Eq:M-tau-tau}) giving $\sin\theta_{13}=0$ now reads
\begin{eqnarray}
M^\prime_{\tau\tau}   = M^\prime_{\mu\mu} + \frac{1-t_{23}^2}{t_{23}} M^\prime_{\mu\tau}.
\label{Eq:M-tau-tau-prime}
\end{eqnarray}

The mass matrix Eq.(\ref{Eq:Mnu}) contains 5 parameters because of the condition of $\textrm{det}(M^{PDG}_\nu)=0$.
Since the Dirac mass matrix $m_D$ is one zero texture, we can analytically express 
the Dirac mass matrix elements in $m_D$ in terms of the light neutrino masses in $M^{PDG}_\nu$.
For instance, if $a_2=0$, the solution consists of 
\begin{eqnarray}
&&{a_1} =  - {\sigma _1}\sqrt {\frac{{{M^\prime_{\mu \mu }}{M^\prime_{ee}} - M_{e\mu }^{\prime 2}}}{{{M^\prime_{\mu \mu }}}}},
\nonumber\\
&&{a_3} =  - {\sigma _3}\sqrt {\frac{{{M^\prime_{\tau \tau }}{M^\prime_{\mu \mu }} - M_{\mu \tau }^{\prime 2}}}{{{M^\prime_{\mu \mu }}}}},
\nonumber\\
&&{b_1} = \frac{{{M^\prime_{e\mu }}}}{{\sqrt {{M^\prime_{\mu \mu }}} }},\quad{b_2} = \sqrt {{M^\prime_{\mu \mu }}},\quad{b_3} = \frac{{{M^\prime_{\mu \tau }}}}{{\sqrt {{M^\prime_{\mu \mu }}} }},
\label{Eq:reconstruction}
\end{eqnarray}
with
\begin{eqnarray}
M^\prime_{e\tau }M^\prime_{\mu \mu } &=&
{\sigma _1}{\sigma _3}\sqrt {\left( {{M^\prime_{ee}}{M^\prime_{\mu \mu }} - M_{e\mu }^{^\prime 2}} \right)\left( {{M^\prime_{\mu \mu }}{M^\prime_{\tau \tau }} - M_{\mu \tau }^{\prime 2}} \right)}
\nonumber\\
&&
+
 {M^\prime_{e\mu }}{M^\prime_{\mu \tau }}, 
\label{Eq:reconstruction_MeeMut}
\end{eqnarray}
due to $\det(M^{PDG}_\nu)=0$, where $\sigma_{1,3}=\pm 1$ \cite {reconstruction, Kitabayashi} and the sign of 
$\sigma_1\sigma_3$ can be calculated. Similarly, other solutions with $a_{1~{\rm or}~3}=0$ or $b_{1, 2~{\rm or}~3}=0$ can be obtained.

According to the condition of $\det(M^{PDG}_\nu)=0$, at least one of the neutrino mass eigenvalues $(m_1,m_2,m_3)$
 must be zero \cite{minimalSeesaw}. We obtain the two types of  hierarchical neutrino mass spectrum in the minimal 
seesaw model. One is the normal mass hierarchy $(m_1,m_2,m_3)=\left(0,\sqrt{\Delta m_\odot^2},\sqrt{\Delta m_{atm}^2}\right)$
 and the other is the inverted mass hierarchy $(m_1,m_2,m_3)=\left(\sqrt{-\Delta m_{atm}^2},\sqrt{\Delta m_\odot^2-\Delta m_{atm}^2},0\right)$.
The matrix elements of $m_D$ can be reconstructed in terms of two mixing angles $\theta_{12,23}$, two neutrino masses $m_{2,3}$, one CP phase $\phi$ and 
two heavy neutrino masses $M_{1,2}$.  

\subsection{Leptogenesis}
In the leptogenesis scenario , the baryon-photon ratio is obtained from the 
baryon asymmetry $Y_B = (n_B-n_{\overline{B}})/s$ as
\begin{eqnarray}
\eta_B = 7.04Y_B,
\end{eqnarray}
where $s$ is entropy density in the universe.  Via the sphaleron process, this baryon asymmetry is related 
to the lepton asymmetry $Y_L$:
\begin{eqnarray}
Y_B= -\frac{8N+4m}{14N+9m} Y_L,
\end{eqnarray}
where $N$ is the number of generation of fermions and $m$  is the number of Higgs doublets. In the 
particle contents of the standard model, we have $Y_B \simeq -0.549Y_L$. 

The lepton asymmetry $Y_L$ is parameterized by three terms as
\begin{eqnarray}
Y_L = d\frac{\epsilon}{g_\ast},
\end{eqnarray}
where $d, \epsilon$ and $g_\ast$ are generally called the dilution factor, CP-asymmetry parameter and 
effective number of the relativistic degree of freedom, respectively.  We use the following estimates: 
\begin{enumerate}
\item 
The dilution factor $d$ should be determined by solving the Boltzmann equation.
 In the present analyses, however, we use the good analytical approximation proposed by Nielsen and 
Takanishi \cite{Nielsen01}:
\begin{eqnarray}
d \sim 
\left\{
 \begin{array}{cc}
    \frac{1}{2.0 \sqrt{K^2+9}} & \quad 0 \le K \le 10 , \\
    \frac{0.3}{K(\ln K)^{0.6}}   &  \quad 10 \le K \le 10^6 ,\\
  \end{array}
\right. 
\end{eqnarray}
where
\begin{eqnarray}
K &=& \frac{M_{pl}}{1.66\sqrt{g_\ast}(8\pi v^2)}\frac{(m_D^\dagger m_D)_{11}}{M_1} \nonumber \\
  &\simeq& \frac{1}{10^{-3} {\rm eV}}\left( \vert a_1 \vert^2+ \vert a_2 \vert^2 + \vert a_3 \vert^2 \right),
\end{eqnarray}
with the Plank mass $M_{pl} \simeq 1.22\times 10^{19}$ GeV and the vacuum expectation value of the 
Higgs field $v \simeq 174$ GeV.
\item 
The CP-asymmetry $\epsilon$ is generated by the decay processes of the heavy neutrinos. If we assume a hierarchical 
mass spectrum of the heavy neutrinos $M_1 \ll M_2$, the interactions of $N_1$ can be in thermal 
equilibrium when $N_2$ decays and the  asymmetry caused by the $N_2$ decay is washed out by the lepton 
number violating processes with $N_1$. Thus, only the decays of $N_1$ are relevant to the generation of 
the final lepton asymmetry. In this case, the CP-asymmetry parameter is calculated to be 
\cite{leptogenesisReviews}
\begin{eqnarray}\label{Eq:epsilon}
&&\epsilon =\frac{M_2}{8\pi v^2}
\frac{\textrm{Im}\left[(a_1^\ast b_1 + a_2^\ast b_2 + a_3^\ast b_3)^2 \right]}{\vert a_1 \vert^2 + 
\vert a_2 \vert^2 + \vert a_3 \vert^2}f\left(\frac{M_2}{M_1} \right),
\label{CP-epsilon}\\
\nonumber
\end{eqnarray}
where the function $f(x)$ is given by
\begin{eqnarray}\label{Eq:loop_function}
&&f(x) = x\left[1 - (1+x^2)\ln\left(\frac{1+x^2}{x^2}\right) + \frac{1}{1-x^2}\right]
\nonumber\\
&&\qquad \approx -\frac{3}{2x}~{\rm for}~x\gg 1,
\end{eqnarray}
\item 
The effective number of the relativistic degree of freedom $g_\ast$ is calculated as \cite{KolbTurner}
\begin{eqnarray}
&&g_\ast = \sum_{i=bosons} g_i\left(\frac{T_i}{T}\right)^4 + \frac{7}{8} \sum_{i=fermions} g_i\left(\frac
{T_i}{T}\right)^4,
\nonumber\\
\end{eqnarray}
where $T$ is thermal equilibrium temperature of the universe, $T_i$ and $g_i$ are temperature and 
number of internal degrees of freedom of the relativistic particle species $i$. For $T \ge 300$ GeV, all 
the species in the standard model are relativistic and we have $g_\ast \simeq 106.75$.
\end{enumerate}

We note that, in the case of $M_1 \ll M_2$, the baryon-photon ratio $\eta_B$ is nearly proportional 
to the mass of heavy neutrino $M_1$ \cite{Kitabayashi}. This $M_1$ dependence to $\eta_B$ can be 
understand by the following rough estimation. For $x=M_2/M_1\gg 1$, 
the CP-asymmetry parameter $\epsilon$ Eq.(\ref{CP-epsilon}) becomes
\begin{eqnarray}
\epsilon \sim \frac{3}{16\pi v^2}M_1\tilde{\epsilon}, 
\quad 
\tilde{\epsilon} \equiv \frac{{\rm Im}[(a_1^\ast b_1 + a_2^\ast b_2 + a_3^\ast b_3)^2]}{\vert a_1 
\vert^2 + \vert a_2 \vert^2 + \vert a_3 \vert^2}.
\label{Eq.epsilon-etaB}
\end{eqnarray}
Since $\tilde{\epsilon}$ is independent of $M_1$, $\epsilon$ is nearly proportional to the mass of 
the heavy neutrino $M_1$. 

\subsection{Numerical analysis}

\begin{figure}[t]
\begin{center}
  \includegraphics{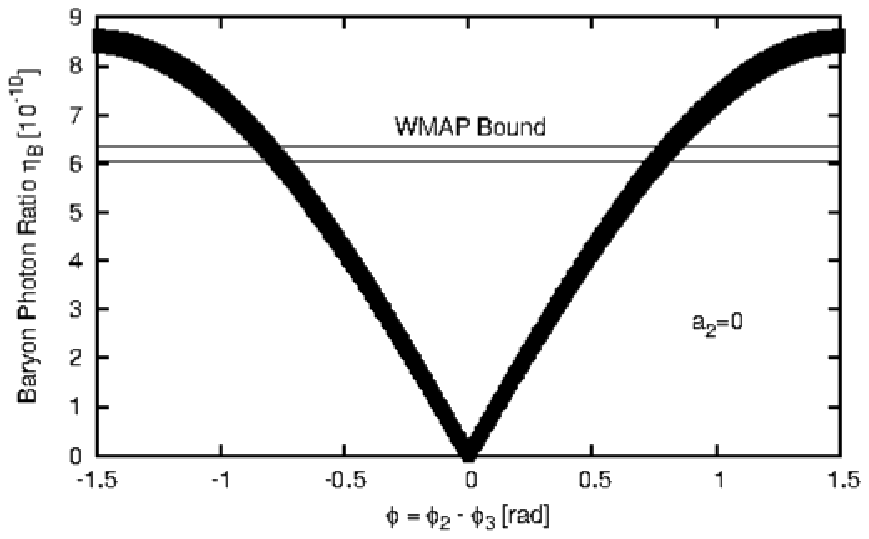}
  \includegraphics{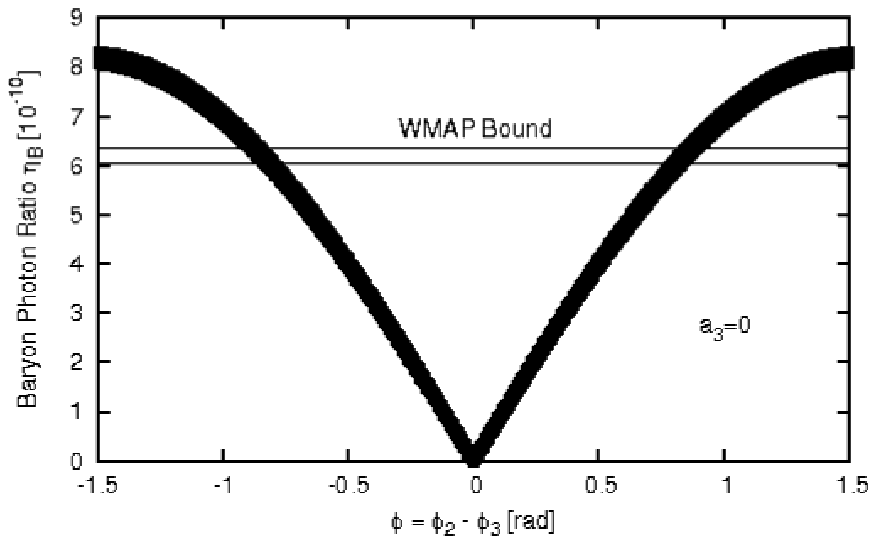}
  \caption{\label{fig:fig1} Majorana CP phase $\phi$ vs the baryon-photon ratio $\eta_B$ for $a_2=0$ (upper figure) and $a_3=0$ (lower figure)
   in the case 1 ($\sin^2\theta_{23} = \tan^2\theta_{12}$) of the bi-pair neutrino mixing.  The horizontal rectangles show the allowed region of $\eta_B$ and the thick black curves show our predictions.}
\end{center}
\end{figure}
\begin{figure}[t]
\begin{center}
  \includegraphics{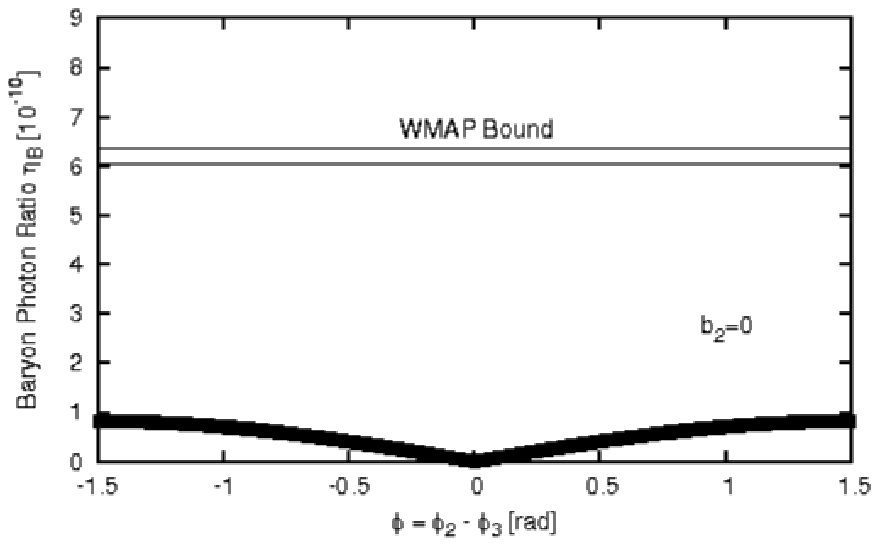}
  \includegraphics{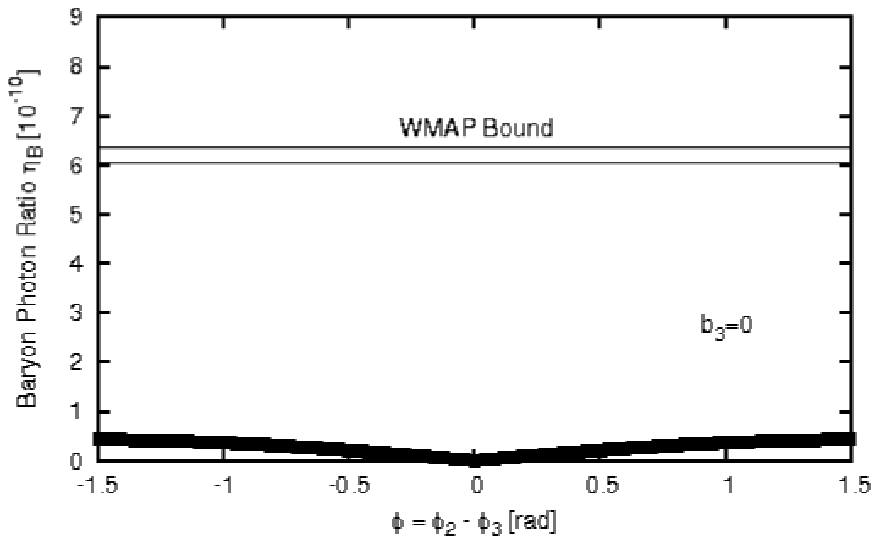}
  \caption{\label{fig:fig2} The same as in FIG.\ref{fig:fig1} but for $b_2=0$ (upper figure) and for $b_3=0$ (lower figure).}
\end{center}
\end{figure}
\begin{figure}[t]
\begin{center}
  \includegraphics{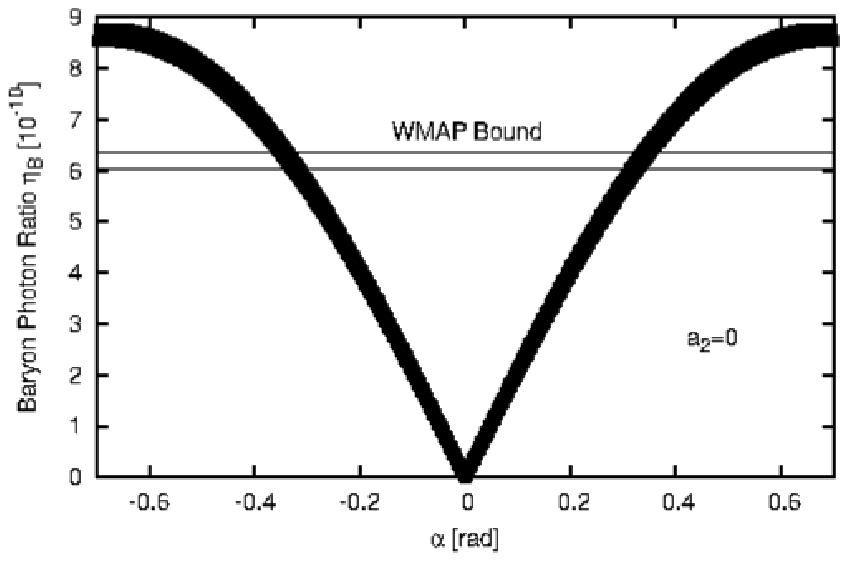}
  \includegraphics{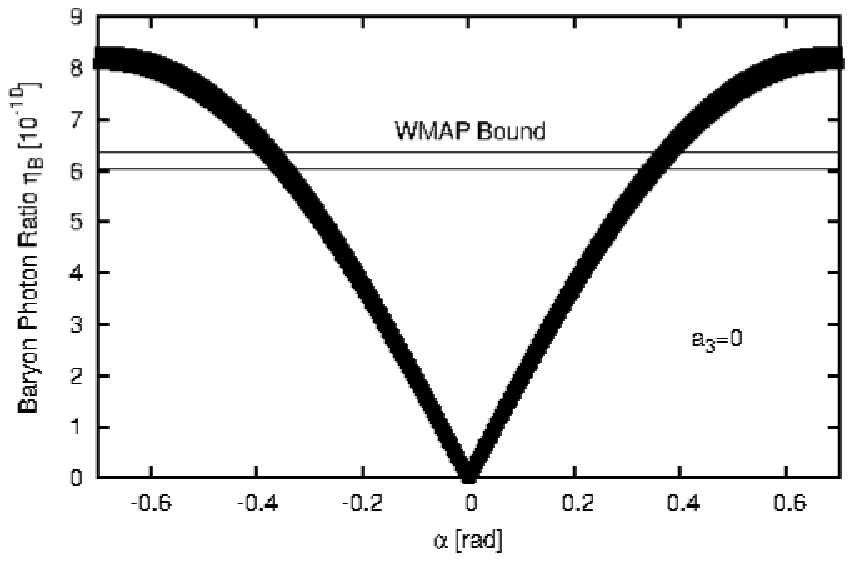}
  \caption{\label{fig:fig3} The same as in FIG.\ref{fig:fig1} but for the phase $\alpha$ for $a_2=0$ (upper figure) and $a_3=0$ (lower figure). }
\end{center}
\end{figure}
\begin{figure}[t]
\begin{center}
  \includegraphics{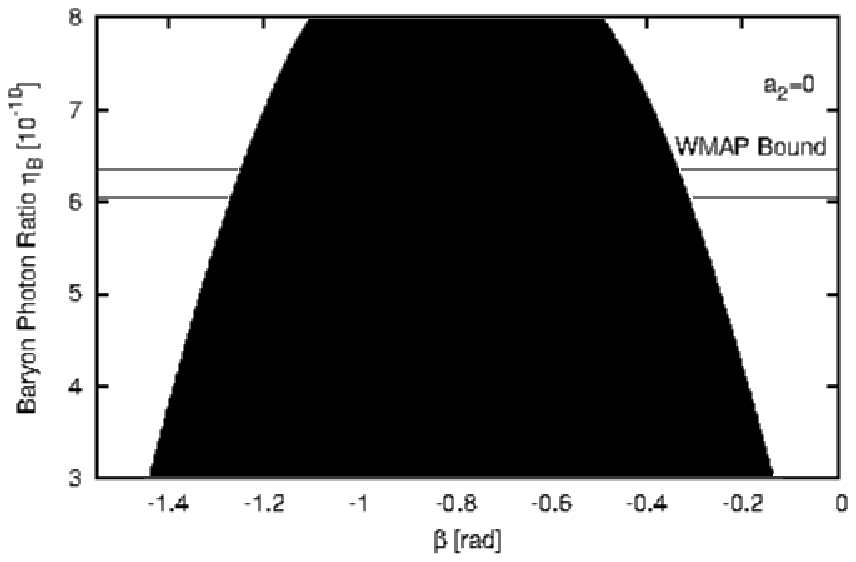}
  \includegraphics{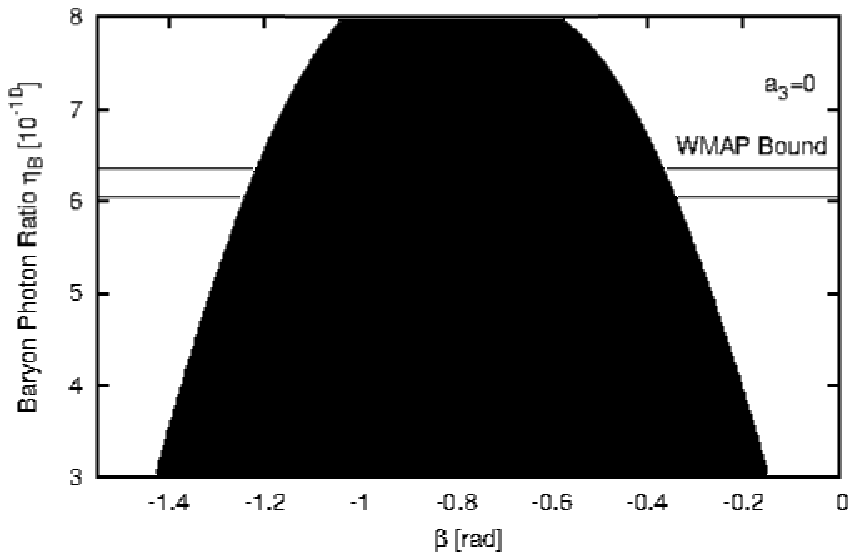}
  \caption{\label{fig:fig4} The same as in FIG.\ref{fig:fig1} but for the phase $\beta$ for $a_2=0$ (upper figure) and $a_3=0$ (lower figure).}
\end{center}
\end{figure}
\begin{figure}[t]
\begin{center}
  \includegraphics{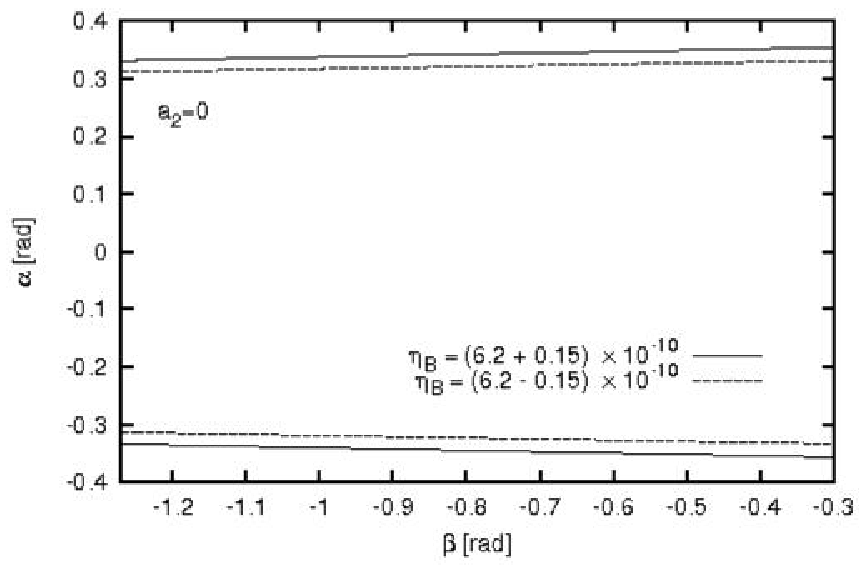}
  \includegraphics{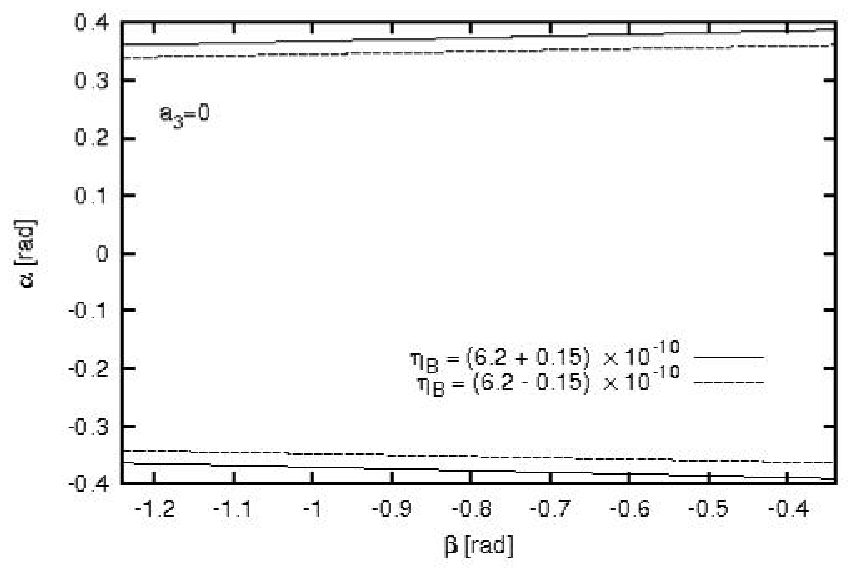}
  \caption{\label{fig:fig5} The allowed regions of the phase $\alpha$ vs the phase $\beta$ indicated by two narrow bands for $a_2=0$ (upper figure) and $a_3=0$ (lower figure)
   in the case 1 ($\sin^2\theta_{23} = \tan^2\theta_{12}$) of the bi-pair neutrino mixing.}
\end{center}
\end{figure}

{\bf Assumptions: }
We have performed the numerical calculation with the following assumptions:  

\begin{enumerate}
\item The light neutrino mass spectrum is the normal mass hierarchy. 
\item $\Delta m_{\odot}^2=7.65 \times 10^{-5} {\rm eV}^2$ and $\left| \Delta m_{atm}^2 \right|=2.40 \times 10^{-3} {\rm eV}^2 $.
\item The heavy neutrino masses $M_{1,2}$ lie between electroweak scale $\sim 10^2$ GeV and GUT scale 
$\sim 10^{15}$ GeV. We take $M_1=5\times 10^{10}(\ll M_2)$ GeV. In order to 
ensure the thermal leptogenesis to be the source of the baryon asymmetry in the universe, the reheating 
temperature after inflation must have been greater than the mass scale of the lightest heavy 
Majorana neutrino \cite{ReheatingTemperature}. Hence the lower bound on the reheating temperature 
must be greater than $\sim 10^{10}$ GeV. However, this high reheating temperature is not suitable for 
supersymmetric (SUSY) theories because it may lead to an overproduction of light supersymmetric 
particles, such a gravitino after inflation \cite{SUSYproblem}. We are not considering this problem 
here and are limiting discussion on non-SUSY cases.
\item The Dirac neutrino mass matrix is one zero texture. However, if either $a_1=0$ or $b_1=0$ is chosen, 
we can prove that $\epsilon \propto a_1^\ast b_1 + a_2^\ast b_2 + a_3^\ast b_3 = 0$ and $\eta_B \propto \epsilon =0$.
The case with $a_1=0$ or $b_1=0$ is excluded.
\item All the particle species in the standard model were relativistic when the leptonic CP-asymmetry 
was generated by the decay process of the lightest heavy neutrino $N_1$. However, $N_1$ was heavy 
enough to be non-relativistic itself. We take $g_\ast=106.75$.
\item We use $\eta_B = (6.2 \pm 0.15) \times 10^{-10}$ as the upper and lower bound of the baryon-photon 
ratio from the WMAP observation \cite{WMAP}.
\end{enumerate}

{\bf Predictions: }
Our results are summarized in five figures FIG.\ref{fig:fig1} $\sim$ FIG.\ref{fig:fig5}.  Basically, since the case with $a_2=0$ ($b_2=0$) 
is identical to the case with $a_3=0$ ($b_2=0$) if the $\mu$-$\tau$ symmetry is exact in $M^{PDG}_\nu$.  The experimental results are consistent with the 
approximate $\mu$-$\tau$ symmetry and we expect that our predictions based on $a_2=0$ ($b_2=0$) are quite similar to those based on $a_3=0$ ($b_3=0$) .
We discuss implications from these figures in the followings:
\begin{enumerate}
\item
{\bf Case 1: }
This case corresponds to the bi-pair neutrino mixing with  $\sin^2\theta_{23} = \tan^2\theta_{12}$.
FIG.\ref{fig:fig1} and FIG.\ref{fig:fig2} show our predictions on $\eta_B$ in the case of $a_2=0$ or $a_3=0$ (FIG.\ref{fig:fig1}) and 
$b_2=0$ or $b_3=0$ (FIG.\ref{fig:fig2}).  
From FIG.\ref{fig:fig1}, we observe that Majorana CP phase $\phi$ lies in the following regions:
\begin{eqnarray}
\vert \phi \vert \sim 0.69 - 0.86 \quad {\rm [rad]},
\end{eqnarray}
for the case of $a_2=0$, and 
\begin{eqnarray}
\vert \phi \vert \sim 0.76 - 0.92 \quad {\rm [rad]},
\end{eqnarray}
for the case of $a_3=0$, where $0 \le \phi \le \pi/2$. 
The differences between the cases of $a_2 = 0$ and of $a_3 = 0$ are not so large (less than 10\%).
On the other hand, FIG.\ref{fig:fig2} shows too small $\eta_B$ and no consistent regions of $\phi$ with the observed $\eta_B$. 
This is naturally expected because $\eta_B$ gets larger for the smaller denominator 
of Eq.(\ref{Eq.epsilon-etaB}) for $\epsilon$, which prefers the case of $a_2=0$ or $a_3=0$ compared with the case of $b_2=0$ or $b_3=0$.

The black regions in FIG.\ref{fig:fig3} and FIG.\ref{fig:fig4} show our predictions of $\alpha$ and $\beta$, which, respectively, stand for the phases of $M_{e\mu}$ and $M_{e\tau}$. 
The horizontal narrow bands in each figures are the allowed regions of the phases, which give
\begin{eqnarray}
&&
0.31 \lesssim \vert \alpha \vert \lesssim 0.37 \quad {\rm [rad]},
\nonumber\\
&&
-1.25 \lesssim \beta \lesssim -0.32 \quad {\rm [rad]},
\end{eqnarray}
for the case of $a_2=0$, and
\begin{eqnarray}
&&
0.34 \lesssim \vert \alpha \vert \lesssim 0.40 \quad {\rm [rad]},
\nonumber\\
&&
-1.23 \lesssim \beta \lesssim -0.35 \quad {\rm [rad]},
\end{eqnarray}
for the case of $a_3=0$. Depicted in FIG.\ref{fig:fig5} is the direct relation of $\alpha$ and $\beta$,  where the range of $\beta$ is restricted to the allowed region. 
From these figures, we observe that 
\begin{itemize}
\item $\alpha$, the phase of $M_{e\mu}$, satisfies  $0.31 \lesssim \vert \alpha\vert \lesssim 0.40$ [rad],
\item $\beta$, the phase of $M_{e\tau}$, lies in the broad range of $-1.25\sim -0.32$ [rad].
\end{itemize}

\item
{\bf Case 2: }
This case corresponds to the bi-pair neutrino mixing with  $\sin^2\theta_{23} = 1-\tan^2\theta_{12}$.
The predictions of $\eta_B$ for $a_2=0$ ($a_3=0$) are reproduced by plots for 
$a_3=0$ ($a_2=0$) in the case 1. 
This correspondence  can be understand by the following way:
The Dirac matrix elements for the case 2 $(a_1^{case 2}, a_2^{case 2}, \cdots)$ can be expressed in terms 
of those for the case 1 $(a_1^{case 1}, a_2^{case 1}, \cdots)$. Namely, we can find that 
\begin{eqnarray}
&& a_1 \vert_{a_2=0}^{case2} = -a_1 \vert_{a_3=0}^{case1},  \quad  b_1 \vert_{a_2=0}^{case2} = -b_1 
\vert_{a_3=0}^{case1}, \nonumber \\
&&  a_2 \vert_{a_2=0}^{case2} = 0, \quad b_2 \vert_{a_2=0}^{case2} =  b_3 \vert_{a_3=0}^{case 1}, \nonumber \\
&& a_3 \vert_{a_2=0}^{case2} =  a_2 \vert_{a_3=0}^{case1},  \quad  b_3 \vert_{a_2=0}^{case2} =  b_2 
\vert_{a_3=0}^{case1},\nonumber \\
&& \sigma_1\sigma_3\vert_{a_2=0}^{case2} = - \sigma_1\sigma_2 \vert_{a_3=0}^{case1}, 
\end{eqnarray}
for $a_2=0$.  As a result, $ a_1b_1^\ast + a_2b_2^\ast + a_3b_3^\ast $ and $ \vert a_1 \vert^2 + \vert a_2 \vert^2 + \vert a_3 \vert^2 $ to 
calculate $\epsilon$ 
 for $a_2=0$ in the case 2 are same as those for $a_3=0$ in the case 1. Therefore, $\eta_B$ 
for $a_2=0$ in the case 2 is equal to $\eta_B$ for $a_3=0$ in the case 1.  Similarly for $a_3=0$ in the case 2.
\end{enumerate}

\section{Summary and Discussions}

We have discussed Majorana CP violation in the recently proposed bi-pair neutrino mixing scheme that predicts
$\sin\theta_{13}=0$ as well as $\tan^22\theta_{12}=2(\sqrt{2}-1)$  with either  $\tan^2\theta_{23}=1/\sqrt{2}$ (the case 1)  
or $\tan^2\theta_{23}=\sqrt{2}$ (the case 2).
Within the minimal seesaw model, where $m_1=0$ is chosen, we have found that the Majorana CP phase $\phi$
is constrained to be:
\begin{eqnarray}
\vert \phi \vert \sim 0.69 - 0.86 \quad {\rm [rad]},
\end{eqnarray}
for the case of $a_2=0$, and 
\begin{eqnarray}
\vert \phi \vert \sim 0.76 - 0.92 \quad {\rm [rad]},
\end{eqnarray}
for the case of $a_3=0$, in order to reproduce the observed WMAP baryon-photon ratio. 

The theoretical study on the flavor structure of the mass matrix giving $\sin\theta_{13}=0$ further reveals that
two Majorana phases $\phi_{2.3}$ are estimated to be:
\begin{eqnarray}
\phi_2&=&-(\alpha+\beta),
\nonumber\\
\phi_3&\approx& \alpha-\beta,
\label{alpha-beta}
\end{eqnarray}
where $\phi=\phi_2-\phi_3\approx -2\alpha$. The phases $\alpha$ and $\beta$, respectively, specify the phases of $M_{e\mu}$ and $M_{e\tau}$ for a given 
neutrino masses $M_{ij}$.  We have estimated that 
\begin{itemize}
\item $\alpha$ satisfies  $0.31 \lesssim \vert \alpha\vert \lesssim 0.40$ [rad],
\item $\beta$ lies in the broad range of $-1.25\sim -0.32$ [rad].
\end{itemize}
The relation Eq.(\ref{alpha-beta}) is based on the calculations of $m_{2,3}$:
\begin{eqnarray}
{m_2}{e^{ - i{\phi _2}}}& =& \frac{{{e^{i\left( {\alpha  + \beta } \right)}}\left| {{M_{e\mu }}} \right|}}{{{c_{12}}{s_{12}}{c_{23}}}},
\nonumber\\
{m_3}{e^{ - i{\phi _3}}}& =& \frac{1}{{s_{23}^2}}\left( {{e^{i\left( {\beta  - \alpha } \right)}}\left| {{M_{\mu \mu }}} \right| - \frac{{{e^{i\left( {\alpha  + \beta } \right)}}{c_{23}}\left| {{M_{e\mu }}} \right|}}{{{t_{12}}}}} \right),
\nonumber\\
\end{eqnarray}
subject to the condition $m^2_3\gg m^2_2$. 
The flavor structure of $M_{ij}$ compatible with the PDG phase convention is given by $M_\nu^{PDG} \vert_{\theta_{13}=0}^{m_1=0}$:
\begin{widetext}
\begin{eqnarray}
M_\nu^{PDG} \vert_{\theta_{13}=0}^{m_1=0} &=&
\left(
  \begin{array}{ccc}
    A e^{i(\alpha + \beta)} \vert M_{e\mu} \vert  
           & e^{i(\alpha + \beta)} \vert M_{e\mu} \vert     
           & -\frac{A}{B} e^{i(\alpha + \beta)} \vert M_{e\mu} \vert 
    \\
           & \frac{A}{t_{12}^2}e^{i(\alpha+\beta)} \vert M_{e\mu} \vert + \frac{A}{B} M_{\mu\tau}  
           & M_{\mu\tau}  
           \\
           &        
           & \frac{A}{t_{12}^2}e^{i(\alpha+\beta)} \vert M_{e\mu} \vert + \frac{B}{A}M_{\mu\tau}
  \end{array}
 \right),
 \end{eqnarray}
\end{widetext}
where $(A, B)$ = $(c_{12}, 1)$ for the case 1 while $(A, B)$ = $(1, c_{12})$ for the case 2.
There exist the relations among masses in $M_\nu^{PDG} \vert_{\theta_{13}=0}^{m_1=0}$, $M^\prime_{ij}$, given by
\begin{eqnarray}
&&\frac{M^\prime_{ee}}{M^\prime_{e\mu}}=A, 
\nonumber\\
&&\frac{M^\prime_{\mu\mu}-\frac{M^\prime_{ee}}{t^2_{12}}}{M^\prime_{\mu\tau}}=\frac{M^\prime_{\mu\tau}}{M^\prime_{\tau\tau}-\frac{M^\prime_{ee}}{t^2_{12}}}=-\frac{M^\prime_{e\tau}}{M^\prime_{e\mu}}=\frac{A}{B},
\end{eqnarray}
which exhibit the characteristic scaling property.  To measure the magnitude of $M^\prime_{ee,e\mu,e\tau}$ in future, we will find that
\begin{eqnarray}
&&
\frac{\vert M_{ee}^\prime\vert}{\vert M_{e\mu }^\prime\vert} = \left\{ \begin{array}{l}
{c_{12}} = \frac{1}{{\sqrt[4]{2}}}  = 0.84 \cdots {\rm{the~case~1}}\\
1 \cdots {\rm{the~case~2}}
\end{array},
\right.
\\
&&
\frac{\vert M_{e\tau }^\prime\vert}{\vert M_{e\mu }^\prime\vert} = \left\{ \begin{array}{l}
{c_{12}} = \frac{1}{{\sqrt[4]{2}}} = 0.84 \cdots {\rm{the~case~1}}\\
\frac{1}{{{c_{12}}}} = \sqrt[4]{2} = 1.19 \cdots {\rm{the~case~2}}
\end{array}, \right.
\end{eqnarray}
if the bi-pair neutrino mixing is correct.

In our future study, we will discuss the detailed feature of the charged lepton corrections arising from the following 
type of a mass matrix:
\begin{eqnarray}
M_\ell =
\left( {\begin{array}{*{20}{c}}
   m_{ee} & \varepsilon_\mu & \varepsilon_\tau  \\
   \varepsilon_\mu & m_{\mu\mu} & m_{\mu\tau}  \\
   \varepsilon_\tau & m_{\mu\tau} & m_{\tau\tau}  \\
\end{array}} \right).
\label{Eq:charged-lepton-mass}
\end{eqnarray}
As effects on the neutrino mixings, the correction to $\theta_{23}$ becomes larger for the larger magnitude of $m_{\mu\tau}$ and 
$\theta_{13}$ becomes nonvanishing for $\varepsilon_{\mu,\tau} \neq 0$.  If there is 
an approximate conservation of the electron number, we may have tiny corrections to $\theta_{13}$ and 
the normal mass hierarchy is welcome to tiny magnitudes of the electron-number-breaking flavor neutrino masses: 
$M_{ee,e\mu, e\tau}$ \cite{e-number}. 
Furthermore, since we have equipped with the general parameterization of $M_\nu$ in 
Eq.(\ref{Eq:Mnu_sin_theta13_0}) compatible with $\sin\theta_{13}=0$, we may discuss Majorana CP violation in more model-independent way or in a way based on 
other specific models giving $\sin\theta_{13}=0$ \cite{Others}.  These subjects will be discussed elsewhere \cite{workInProgress}.



\begin{thebibliography}{99}

\bibitem{atmospheric}
Y. Fukuda, et al., Super-Kamiokande Collaboration, \Journal{\PRL}{81}{1562}{1998}; \Journal{\PRL}{82}{2430}{1999};
T. Kajita, \Journal{\NPBSUPPL}{77}{123}{1999};
See also, T. Kajita and Y. Totsuka, \Journal{\RMP}{73}{85}{2001}.

\bibitem{solar}
J. N. Bahcall, W. A. Fowler, I. Iben and R. L. Sears, \Journal{\APJ}{137}{344}{1963};
J. Bahcall, \Journal{\PRL}{12}{300}{1964};
R. Davis Jr., \Journal{\PRL}{12}{303}{1964};
R. Davis Jr., D. S. Harmer and K. C. Hoffman, \Journal{\PRL}{20}{1205}{1968};
J. N. Bahcall, N. A. Bahcall and G. Shaviv, \Journal{\PRL}{20}{1209}{1968};
J. N. Bahcall and R. Davis Jr., \Journal{\SCIENCE}{191}{264}{1976};
Y. Fukuda, et al., Super-Kamiokande Collaboration, \Journal{\PRL}{81}{1158}{1998}; \Journal{\PRL}{81}{4279}{1998}, Erratum;
B. T. Clevel, et al., Super-Kamiokande Collaboration, \Journal{\APJ}{496}{505}{1998};
W. Hampel, et al., GNO Collaboration, \Journal{\PLB}{447}{127}{1999};
Q. A. Ahmed, et al., SNO Collaboration, \Journal{\PRL}{87}{071301}{2001}; \Journal{\PRL}{89}{011301}{2002}.

\bibitem{reactor}
See for example,
K. Eguchi, et al., KamLAND Collaboration, \Journal{\PRL}{90}{021802}{2003};
S. Abe, et al., KamLAND Collaboration, \Journal{\PRL}{100}{221803}{2008}.

\bibitem{theta13}
M. Apollonio et al., CHOOZ Collaboration, \Journal{\EPJC}{27}{331}{2003};
X. Guo et al., Daya-Bay Collaboration, {\it A precision measurement of the neutrino mixing angle 
$\theta_{13}$ using reactor antineutrinos at Daya Bay}, arXiv:hep-ex/0701029; 
F. Ardellier et al., Double Chooz Collaboration, {\it Double Chooz: A search for the neutrino mixing 
angle $\theta_{13}$}, arXiv:hep-ex/0606025; 
C. Palomares, {\it Double-Chooz Neutrino Experiment}, arXiv:0911.3227 [hep-ex]; 
J. K. Ahn et al., RENO Collaboration, {\it RENO: An Experiment for Neutrino Oscillation Parameter 
$\theta_{13}$ Using Reactor Neutrinos at Yongg- wang}, arXiv:1003.1391 [hep-ex].

\bibitem{accelerator}
See for example,
S. H. Ahn, et al., K2K Collaboration, \Journal{\PLB}{511}{178}{2001}; \Journal{\PRL}{90}{041801}{2003}.

\bibitem{Xing2010}
For a review, see for example, Z.-Z. Xing, \Journal{\NPBSUPPL}{203-204}{82}{2010}.

\bibitem{Schwetz2008}
T. Schwetz, M. T\'{o}rtola and J.W.F. Valle, \Journal{\NJP}{10}{113011}{2008};
G. L. Fogli, E. Lisi, A. Marrone, A. Palazzo and A.M. Rotunno, \Journal{\PRL}{101}{141801}{2008}.

\bibitem{UMNS}
B. Pontecorvo, \Journal{\JETPUSSR}{34}{247}{1958};
Z. Maki, M. Nakagawa and S. Sakata, \Journal{\PTP}{28}{870}{1962}.

\bibitem{MajoranaPhase}
S.M. Bilenky, J. Hosek and S.T. Petcov, \Journal{\PLBOLD}{94B}{495}{1980};
J. Schechter and J.W.F. Valle, \Journal{\PRD}{22}{2227}{1980};
M. Doi, T. Kotani, H. Nishiura, K. Okuda and E. Takasugi, \Journal{\PLBOLD}{102B}{323}{1981}.

\bibitem{T2K}
K. Abe, et. al., T2K Collaboration, \Journal{\PRL}{107}{041801}{2011};
Y. Oyama, T2K Collaboration, {\it Current status of the T2K experiment}, arXiv:1108.4457 [hep-ex].

\bibitem{MINOS}
P. Adamson {\it et. al.}, MINOS Collaboration, \Journal{\PRD}{82}{051102}{2010};
L. Whitehead, et. al., MINOS Collaboration, {\it Recent result from MINOS}, Joint Experimental-Theoretical 
Seminar (24 June 2011, Fermilab, USA). 
See also, G. L. Fogli, E. Lisi. A. Marrone, A. Palazzo and A. M. Rotunno, {\it Evidence of $\theta_{13} > 0$
 from global neutrino data analysis}, arXiv:1106.6028 [hep-ph].

\bibitem{Mohapatra_Pal}
See for review, R. N. Mohapatra and P. B. Pal, {\it Massive Neutrinos in Physics and Astrophysics}, 3rd. ed. World Scientific (2004).

\bibitem{DoubleBeta}
See for example.
D. Delepine, V.G. Macias, S. Khalil and G.L. Castro, \Journal{\PLB}{693}{438}{2010};
A. Joniec, \Journal{\ACTAPPB}{37}{2171}{2006}.

\bibitem{leptogenesisReviews}
M. Fukugita and T. Yanagida, \Journal{\PLB}{174}{45}{1986};
J. A. Harvey and M. S. Turner, \Journal{\PRD}{42}{3344}{1990};
M. A. Luty, \Journal{\PRD}{45}{455}{1992};
M. Pl\"{u}macher, \Journal{\ZPC}{74}{549}{1997};
W. Buchm\"{u}ller and M. Pl\"{u}macher, \Journal{\PLB}{431}{354}{1998}.
 
\bibitem{WMAP}
E. Komatsu et. al., \Journal{\APJS}{180}{330}{2009}.

\bibitem{mu-tau}
T. Fukuyama and H. Nishiura, in {\it Proceedings of International Workshop on Masses and Mixings of Quarks 
and Leptons}, Shizuoka, 1997, edited by Y. Koide (World Scientific, Singapore, 1997), p.252; {\it Mass 
Matrix of Majorana Neutrinos}, arXiv:hep-ph/9702253;
R. N. Mohapatra and S. Nussinov, \Journal{\PRD}{60}{013002}{1999};
E. Ma and M. Raidal, \Journal{\PRL}{87}{011802}{2001};
C. S. Lam, \Journal{\PLB}{507}{214}{2001};
P. F. Harrison, W. G. Scott, \Journal{\PLB}{574}{219}{2002};
W. Grimus and L. Lavoura, \Journal{\PLB}{572}{189}{2003};
  \Journal{\JPG}{30}{73}{2004};
Y. Koide, \Journal{\PRD}{69}{093001}{2004};  
T. Ota and W. Rodejohann, \Journal{\PLB}{639}{322}{2006}.

\bibitem{KitabayashiYasue}
T. Kitabayashi and M. Yasu\`{e}, \Journal{\PLB}{524}{308}{2002};
  \Journal{\IJMPA}{17}{2519}{2002};
  \Journal{\PRD}{67}{015006}{2003};
  \Journal{\PLB}{621}{133}{2005};
I. Aizawa, M. Ishiguro, T. Kitabayashi and M. Yasu\`{e}, \Journal{\PRD}{70}{015011}{2004}; 
I. Aizawa, T. Kitabayashi and M. Yasu\`{e},  \Journal{\PRD}{71}{075011}{2005};
 \Journal{\NPB}{728}{220}{2005}. 

\bibitem{Tribimaximal}
P.F. Harrison, D.H. Perkins and W.G. Scott, \Journal{\PLB}{530}{167}{2002};
Z.-Z. Xing, \Journal{\PLB}{533}{85}{2002};
P.F. Harrison and W.G. Scott, \Journal{\PLB}{535}{163}{2002}.

\bibitem{Scaling}
R. N. Mohapatra and W. Rodejohann, \Journal{\PLB}{644}{59}{2007};
A. Blum, R. N. Mohapatra and W. Rodejohann, \Journal{\PRD}{76}{053003}{2007}.

\bibitem{Others}
For other parameterizations, 
see H. Fritzsch and Z.-Z. Xing, \Journal{\PLB}{372}{265}{1996};
V.D. Barger, S. Pakvasa, T.J. Weiler and K. Whisnant, \Journal{\PLB}{437}{107}{1998};
C. Giunti, \Journal{\NPBSUPPL}{117}{24}{2003}; 
Z. -Z. Xing, \Journal{\JPG}{29}{2227}{2003};
Y. Kajiyama, M. Raidal, A. Strumia, Phys. Rev. D 76, \Journal{\PRD}{76}{117301}{2007};
W. Rodejohann, \Journal{\PLB}{671}{267}{2009};
C. H. Albright, A. Dueck and W. Rodejohann, \Journal{\EPJC}{70}{1099}{2010};
G.-J. Dinga, L.L. Everett and A.J. Stuart, {\it Golden Ratio Neutrino Mixing and A5 Flavor Symmetry}, 
arXiv:1110.1688 [hep-ph].
See also, I. de Medeiros Varzielas, R. Gonz\'{a}lez Felipe and H. Ser\^{o}dio, \Journal{\PRD}
{83}{033007}{2011}. 

\bibitem{Bipair}
T. Kitabayashi and M. Yasu\`{e}, \Journal{\PLB}{696}{478}{2011}.

\bibitem{Seesaw}
P. Minkowski, \Journal{\PLB}{67}{421}{1977}; 
T. Yanagida, in  {\it Proceedings of the Workshop on the Unified Theory and Baryon 
Number in the Universe}, KEK, 1979, edited by O. Sawada and A. Sugamoto (KEK report 79-18, 1979), p.95; 
\Journal{\PTP}{64}{1870}{1980}; 
M. Gell-Mann, P. Ramond and R. Slansky, in {\it Supergravity, Proceedings of the Supergravity Workshop}, 
Stony Brook, 1979, edited by P. van Nieuwenhuizen and D.Z. Freedmann (North-Holland, Amsterdam 1979), p.315;
R.N. Mohapatra and G. Senjanovi\`{c}, \Journal{\PRL}{44}{912}{1980}. 
See also, P. Minkowski, in {\it Proceedings of the X\hspace{-.1em}I International
Workshop on Neutrino Telescopes in Venice}, Venice, 2005, edited by M. Baldo Ceolin 
(Papergraf S.p.A, Italy, 2005), p.7.

\bibitem{minimalSeesaw}
T. Endoh, S. Kaneko, S.K. Kang, T. Morozumi and M. Tanimoto, \Journal{\PRL}{89}{231601}{2002};
P.H. Frampton, S.L. Glashow and T. Yanagida, \Journal{\PLB}{548}{119}{2002};
M. Raidal and A. Strumia, \Journal{\PLB}{553}{72}{2003};
R.G. Felipe, F.R. Joaquim and B.M. Nobre, \Journal{\PRD}{70}{085009}{2004}.

\bibitem{reconstruction}
V. Barger, D. A. Dicus, H-J. He and T. Li, \Journal{\PLB}{583}{173}{2004};
S. Chang, S. K. Kang and K. Siyeon, \Journal{\PLB}{597}{78}{2004}.

\bibitem{PDG}
K. Nakamura, et.al., Particle Data Group, \Journal{\JPG}{37}{075021}{2010}.

\bibitem{BabaAndYasue}
T. Baba and M. Yasu\`{e}, \Journal{\PRD}{75}{055001}{2007}; \Journal{\PRD}{77}{075008}{2008};
\Journal{\PTP}{123}{659}{2010}.
See also,
S.-F. Ge, H.-J. He and F.-R. Yin,  \Journal{\JCAP}{1005}{017}{2010};
Z.-Z. Xing and Y.-L. Zhou, \Journal{\PLB}{693}{584}{2010}.

\bibitem{YudaAndYasue}
K. Yuda and M. Yasu\`{e}, \Journal{\PLB}{693}{571}{2010}.

\bibitem{Kitabayashi}
T. Kitabayashi, \Journal{\PRD}{76}{033002}{2007}; \Journal{\PTP}{120}{443}{2008}.

\bibitem{Nielsen01}
H. B. Nielsen and Y. Takanishi, \Journal{\PLB}{507}{241}{2001}.

\bibitem{KolbTurner}
E. W. Kolb and M. S. Turner, {\it The Early Universe}, Addison-Wesley (1990).

\bibitem{ReheatingTemperature}
S. Davidson and A. Ibarra, \Journal{\PLB}{535}{25}{2002};
K. Hamaguchi, H. Murayama and T. Yanagida, \Journal{\PRD}{65}{043512}{2002};
W. Buchm\"{u}ller, P. Di Bari and M. Pl\"{u}macher, \Journal{\NPB}{665}{445}{2003};
T. Hambye, Y. Lim, A. Notari, M. Papucci and A. Strumia, \Journal{\NPB}{695}{169}{2004};
G. F. Giudice, A. Notari, M. Raidal, A. Riotto and A. Stumia, \Journal{\NPB}{685}{89}{2004};
W. Buchm\"{u}ller, P. Di Bari and M. Pl\"{u}macher, \Journal{\ANP}{315}{305}{2005}.

\bibitem{SUSYproblem}
See for example, 
M.Yu. Khlopov and A.D. Linde, \Journal{\PLBOLD}{138B}{265}{1984};
F. Balestra, G. Piragino, D.B. Pontecorvo and M.G. Sapozhnikov, I.V.Falomkin and M.Yu.Khlopov,
\Journal{\YFIZ}{39}{990}{1984} [\Journal{\SJNP}{39}{626}{1984}];
M.Yu. Khlopov, Yu.L. Levitan, E.V. Sedelnikov and I.M. Sobol,
\Journal{\YFIZ}{57}{1466}{1994} [\Journal{\PAN}{57}{1393}{1994}].
See also,
W. Buchm\"{u}ller, R. D. Peccei and T. Yanagida, 
{\it Leptogenesis as the origin of matter}, arXiv: hep-ph/0502169.

\bibitem{e-number}
I. Aizawa, M. Ishiguro, T. Kitabayashi and M. Yasu\`{e}, in Ref.\cite{KitabayashiYasue}. 

\bibitem{workInProgress}
T. Kitabayashi and M. Yasu\`{e}, work in progress.

\end{thebibliography}
\end{document}